\documentclass[a4paper]{jpconf}
\usepackage{graphicx}
\begin{document}
\title{Challenge of polarized beams at future colliders}

\author{Gudrid Moortgat-Pick$^{1,2}$,
I.R.\ Bailey$^{3,2}$, D.P. Barber$^{4,3,2}$, E.\ Baynham$^{5}$, A. Birch$^{3,2}$,  T.\ Bradshaw$^{5}$, A.\ Brummitt$^{5}$, S.\ Carr$^{5}$,
J.A.\ Clarke$^{6,2}$, P.\ Cooke$^{3,2}$, J.B.\ Dainton$^{3,2}$, T.~Hartin$^{7}$,  Y.\ Ivanyushenkov$^{5}$, L.J.\ Jenner$^{3,2}$, A.\ Lintern$^{5}$,
L.I.\ Malysheva$^{3,2}$,  O.B.\ Malyshev$^{6,2}$, J.\ Rochford$^{5}$,
 S. Riemann$^{4}$, A. Sch\"alicke$^{4}$, P.~Schmid$^{4}$,  D.J.\ Scott$^{6,2}$, A. Ushakov$^{4}$, 
 Lei Zang$^{3,2}$\\
}

\address{ $^{1}$ Institute for Particle Physics Phenomenology, University of Durham, Durham DH1 3LE, UK\\ 
$^{2}$ Cockcroft Institute,
Daresbury Laboratory, Warrington, Cheshire WA4 4AD, U.K.\\ 
$^{3}$
Department of Physics, University of Liverpool, Oxford St., Liverpool,
L69 7ZE, U.K.\\ 
$^{4}$ DESY, Deutsches Elektronen Synchrotron,
Notkestra{\ss}e 85, D-22607 Hamburg, Germany\\
 $^{5}$ CCLRC
Rutherford Appleton Laboratory, Chilton, Didcot, Oxfordshire OX11 0QX,
U.K.\\
$^{6}$ CCLRC ASTeC Daresbury Laboratory,
Daresbury, Warrington, Cheshire WA4 4AD, U.K.\\
$^{7}$ John Adams
Institute, Oxford University Physics, Oxford OX1 3RH, U.K.
}

\ead{g.a.moortgat-pick@durham.ac.uk}

\begin{abstract}
A short overview about the potential of polarized beams at future colliders is given.
In particular the baseline design for polarized beams at the ILC is 
presented and the physics case for polarized $e^-$ and $e^+$ is discussed.
In order to fulfil the precision requirements spin tracking 
from the source to the interaction point is needed. 
Updates concerning the theoretical calculations as well as their implementation in simulation codes
are reported.  
\end{abstract}

\vspace{-1cm}
\section{Introduction}
\subsection{Overview about colliders with polarized beams}
As shown in past particle physics experiments,  beam polarization is a very
powerful tool to achieve physics goals and optimize results. The great success of the
SLD experiments at the $e^+e^-$ collider SLC with the best 
single measurement of the electroweak mixing angle, 
$\sin^2\theta_{\rm eff}=0.23098\pm 0.00026$~\cite{sld}, 
was due to the application of 
polarized $e^-$ beams with about $P_{e^-}=78\%$.
Although the LEP $e^+e^-$ experiment  had much higher luminosity,
a larger statistical uncertainty, $\sin^2\theta_{\rm eff}=0.23221\pm 0.00029$, was derived.
The polarization of the beams at LEP caused by the Sokolov-Ternov effect
was very small and could not be exploited for physics analyses, but was
nevertheless very useful for calibrating the energy of the beams~\cite{lep}.
Polarization at HERA, the asymmetric circular $ep$-collider, 
reached a polarization of $P_{e^\pm}=40\%$ to $50\%$ at low background 
in the colliding mode (about 70\% in non-colliding mode)
and was used
to test the non-existence of right-handed charged currents~\cite{hera}.

Many of the designs of future colliders also foresee the option
of polarized beams. For instance, future upgrades of $ep$ colliders, i.e.\ 
eRHIC~\cite{erhic} and LHeC~\cite{lhec}, may include  
this option in order to have access to the 
spin structure of the gluons.
 The most prominent future collider with polarized beams is the 
$e^+e^-$ International Linear Collider (ILC), which is already in the engineering phase of its design. 
An electron beam with a polarization between $80\%$ and $90\%$ is included in the baseline design: using
the same scheme for producing polarized electrons as was already successfully demonstrated at the SLC.
Furthermore the baseline $e^+$ source, based on undulator radiation~\cite{balakin79}, 
generates polarized $e^+$ with high luminosity and a predicted 
polarization of about 30\%. The degree of the polarization can easily be 
upgraded to about 60\%~\cite{rdr}.
The option of using polarized $e^-$ and $e^+$ beams at CLIC, 
a future
multi-TeV collider, is also being considered as part of the current design studies~\cite{posipol07}.

\subsection{Physics motivation for polarized beams at the ILC}
Polarizing both beams at the linear collider instead of only the $e^-$
beam has several advantages: improving statistics, enhancing rates and
cross sections and suppressing background processes. 
Furthermore there exist several examples were having both beams 
polarized is mandatory, for instance, in order to determine specific  
quantum numbers of new particles. The polarization of both beams is also 
needed to achieve the ultimate precision predicted for the
measurements at GigaZ.
The physics case and the need of 
polarized $e^-$ and $e^+$ has been 
established and quantified~\cite{Moortgat:2005cw}. 

A striking feature of the current ILC design is that it  provides
without any upgrades
an $e^+$ polarization of about 30\%. Numerous questions could
already be addressed with such an amount of polarization. In many
cases polarized beams with $(P_{e^-},P_{e^+})=(80\%,30\%)$ lead to already half of the
physics gain that could be achieved with $(80\%,60\%)$. These gains
could not be achieved by using higher electron polarization alone,  not even with
100\% $e^-$ beam polarization~\cite{Moortgat:2005cw,Riemann}.

\section{Schemes for polarizing beams at the linear collider}
The electron source consists of a circularly polarized high-power
laser beam and a high-voltage DC gun with a semiconductor
photocathode.  
For the positron source a scheme, based on helical-undulator radiation,
has been chosen as the most reliable solution for producing the required flux of order
$10^{14}$ positrons per pulse (for details see~\cite{rdr,source}). 
The design produces positrons via an
electromagnetic shower instigated in a thin target by incident
circularly polarized synchrotron radiation produced by the undulator
operating on the main ILC $e^-$ beam.  The undulator-based source
produces 1.5 positrons per an electron in the main linac as required to guarantee smooth operation of the ILC,
 and imposes much less demands for
capture issues and damping ring acceptance than conventional technologies. This
method has been
experimentally tested in the E166 experiment at SLAC~\cite{e166}
and 
several prototypes for 
the ILC-type undulator have already been successfully
tested (for details see~\cite{Scott:2007zza}). 
Studies and simulations show that the undulator-based source has negligible 
impact on the emittance and on the
energy spread of the $e^-$ beam~\cite{Scott:2007zza}.
 The undulator-based $e^+$ source leads to much less
radiation damage at the target: for instance, it causes less activation
(dose rate) by a factor of about 70 (25) and produces less neutrons by
about a factor of 10 compared with the target at a conventional
source~\cite{Ushakov:2006th}.  Concerning the status of prototype
targets for the ILC, see~\cite{Bailey:2006zza}. 

The successful accomplishment of the experiment E166 led to the
inclusion of polarization in the physics simulation program
GEANT4~\cite{Dollan:2005nj}. This is important for physics analyses at
all future colliders, and an updated version of the program is now
publically available~\cite{Geant4}. This polarization extension is now
being used in several simulation studies around the polarized positron
source, for instance, for the design and optimization of a low-energy
positron polarimeter~\cite{lepol}.

An alternative scheme for the inclusion of polarized $e^+$ beams at
the linear collider is based on laser-Compton-backscattering. Prototypes for
this scheme have been successfully tested at
ATF~\cite{Omori:2005eb}. Several applications of this scheme to future accelerators
have been discussed including
SuperB factories, a possible multi-TeV design for a future linear
collider CLIC, and energy-recovery linacs (ERL).

\section{Spin tracking from source to the interaction point}
It is important to ensure that no significant polarization is lost 
during the transport of the $e^-$ and $e^+$ beams from their sources to 
the interaction region. The largest effects are expected to be caused by the 
collision of the two beams at the interaction point~\cite{thompson}.
Transport elements downstream of the
sources which can contribute to a loss of polarization include the
initial acceleration structures, transport lines to the damping rings,
the damping rings, the spin rotators~\cite{schmid}, the main linacs, and the high
energy beam delivery systems; as overview, for instance, see~\cite{Smith:2007zz}.
\subsection{Beam-beam interactions}
The main sources of depolarization effects during beam-beam 
interactions are the spin precession 
and the spin-flip processes, i.e.\ the Sokolov-Ternov (S-T) effect.
 Usually the spin precession 
effect is dominant, but at higher energy
the depolarization due to the S-T effect increases~\cite{eurotev}.
Spin precession is described by the Thomas--Bargman-Michel-Telegdi (T-BMT) 
equation, 
\begin{equation}
\frac{{\rm d}\vec{S}}{dt}=-\frac{e}{m\gamma}[(\gamma a+1) 
\vec{B}_T+(a+1)\vec{B}_L-\gamma(a+\frac{1}{\gamma+1})\beta \vec{e}_v\times \frac{\vec{E}}{c}]\times \vec{S},
\label{tbmt}
\end{equation}
where $a$ describes the anomalous magnetic moment of the electron
given by the higher-order corrections to the $ee\gamma$ vertex.  In
the environment of strong colliding beams, however, the usual
perturbation theory cannot be applied.  Therefore modified expressions
for the anomalous magnetic moment in a medium have been
derived~\cite{Baier:2000yv}. These expressions have been evaluated in
the no-scattering case, using the quasi-classical approximation that
implies that the momentum change due to the strong fields 
occurs slowly on the scale of the particle wavelength.  This condition is fulfilled if the Larmor
radius of the particle due to the existing magnetic field in the
bunches is much larger than the particle wavelength. It has been
checked that even in the strong field environment of the ILC such a
quasi-classical approximation can be used and the modified T-BMT equation can be
applied to describe the spin precession sufficiently accurately, see
also~\cite{Baier:2000yv,cain,soon}.

The production of incoherent background pairs~\cite{Rimbault:2006ik} 
is strongly dependent on the
polarization state of the initial photons involved in the
process~\cite{Hartin:2007zza}. These photons are either real (beamstrahlung) or virtual and
depend on the electromagnetic field of the oncoming beam.  The
CAIN~\cite{cain} program contained only full polarizations for the real
photons. The polarization of virtual photons depends on the beam
electric field $E_\omega^{x,y}$ at the point $(x,y)$ where the pair is
produced. For gaussian bunches an analytical expression has been
derived~\cite{engel} and can be solved by using the condition for flat
beams $\sigma_x\gg\sigma_y$.  The cross-section for the Breit-Wheeler process  is also required with
full polarizations. In CAIN this
cross section $\sigma^{\rm circ}$ was written down only for the
product of circular polarizations $\xi_2\xi'_2$ of initial photons $k$
and $k'$. The full cross-section $\sigma^{\rm full}$ is a sum over
all polarization states and functions of final electron energy
$\epsilon$ and momentum $p$ \cite{baier}. A numerical investigation of
these two cross-sections reveals that the usual peak at low energies 
is substantially reduced when using the full cross-section 
for electron
energies less than approximately 50 MeV.  CAIN was modified with the
above expressions and was run for all seven 500 GeV centre of mass
collider parameter sets, cf. also~\cite{eurotev}. There was a $10\%$ to $20\%$
overall reduction in pairs,
with no discernible impact on collision luminosity~\cite{Hartin:2007zza}.

The coherent production of pairs via the first order interaction
between beamstrahlung photon and beam field is  already included in
CAIN. However the second order stimulated Breit-Wheeler process also
takes place in the presence of the bunch fields. The cross-section
calculation involves solutions of the Dirac equation in an external
field. Naively, in comparison to the first order coherent process, the
second-order cross-section should be diminished by an order of the fine
structure constant. However the bunch field has the effect of allowing
the second order cross-section to reach the on-shell mass.  The resulting
resonances are rendered finite by inclusion of the electron
self-energy and the stimulated Breit-Wheeler cross-section can exceed
the first order coherent process.  A detailed theoretical and
numerical investigation is required to gauge in detail the effect on produced
pairs \cite{soon}.
 
%
\subsection{Spin transport}
The
SLICKTRACK~\cite{slicktrack} Monte Carlo computer code has been used to
analyze the spin motion in the ILC damping ring (DR), main linac (ML) and
beam delivery system (BDS).  The simulation, applied for the 6km DR lattice
at 5.0~GeV, shows that the sum of the mean squares of the angles of the tilts
of spins away from the direction of the equilibrium polarization will
be less than 0.1 mrad$^2$, even after 8 damping times. Also close to
the spin-orbit resonance at 4.8~GeV the sum of the mean squares of the
angles is only about 40~mrad$^2$, i.e. still negligible. In case a
large energy spread was included in the simulation of about $\pm
25$~MeV, much greater than the natural energy spread of the DR, the
deviation was found to be 20~mrad$^2$, which is once again negligible~\cite{Malysheva:2007zz}.

A striking result is that the horizontal projections of the spin
vectors of an $e^-$ or $e^+$ bunch do not fully decohere, even after
8000 turns. In other words, if the spins are not perfectly oriented
vertically at injection then their projections do not fan out uniformly
in the horizontal plane during the damping~\cite{Malysheva:2007zz}.
 SLICKTRACK has also been modified to simulate spin tracking in
the ILC beam delivery system with an 2mrad crossing angle, including
realistic misalignments. Consistent with~\cite{Smith:2007zz} it was
found that a depolarization of $<0.06\%$ can be expected.
Since the main linac in the current ILC design
follows the Earth curvature, a spin
precession of about 26 degrees is expected and the ratio between
final and initial polarizations is about $\cos(10^{-4}$rad)~\cite{Malysheva:2007zz}.

\section{Conclusions}
Polarized beams are required to achieve many physics goals and to maximize the number of possible measurements at a collider facility,
 and are a basic ingredient
for many present and future accelerator designs.  The ILC provides
already in the baseline a polarized $e^-$ source with about
$P_{e^-}=80\%-90\%$ and a polarized $e^+$ beam with about $P_{e^+}\sim 30\%$, extendable
to at least $P_{e^+}\le 60\%$, using undulator radiation to produce positrons. The
scheme has been successfully tested at the E166 experiment and several 
undulator prototypes have been accomplished that already achieve the
ILC requirements.  

Precise spin tracking is a necessary condition for
successfully applying polarized beams for physics. Much progress has
been made describing spin motion during beam-beam interactions,
in the damping ring, in the main linac and in 
the beam delivery system. 
Theoretical updates of the used calculations and the description of coherent and incoherent background
processes including  
higher-order contributions have been accomplished. The analytically-based 
program CAIN has also been correspondingly updated.

With the code SLICKTRACK several simulations for
different ILC lattices have been performed. They showed that only small
depolarization can be expected at the ML. The depolarization in the BDS is small but not negligible.
No full decoherence of horizontally spin components in the DR, however,  can be expected. 
Proper alignment of the positron's spin directions prior to injection into the damping rings
even for nominally unpolarized beams is therefore needed.

In order to guarantee that the produced polarization can be
successfully utilized for physics analyses accurate polarimeters are
needed. It has not yet been determined at what frequency
the helicities of the beams have to be flipped between the possible polarization
configurations in order to control systematic uncertainties on conditions at the IP. 
However, in order to fulfil both the high-luminosity as well as the high 
precision goals for physics analyses at the ILC, flipping of the
helicities of the $e^-$ beams as well as the $e^+$ beam is absolutely 
needed~\cite{Riemann}.
News on the positron source engineering design for the ILC and further polarization issues
can be obtained from the working group of the 
ILC positron source group, see also~\cite{source}.

\end{document}